\documentstyle[psfig]{mn0}

\def\simg{\mathrel{\rlap{\raise 0.511ex \hbox{$>$}}{\lower 0.511ex \hbox{$\sim$}}}}
\def\siml{\mathrel{\rlap{\raise 0.511ex \hbox{$<$}}{\lower 0.511ex \hbox{$\sim$}}}}
\def\beq{\begin{equation}} \def\eeq{\end{equation}}

\begin{document}

\title [X-ray afterglows from scattered emission]
  {X-ray flares, plateaus, and chromatic breaks of GRB afterglows from up-scattered
  forward-shock emission}

\author[A. Panaitescu]{A. Panaitescu \\
       Space Science and Applications, MS D466, Los Alamos National Laboratory, 
       Los Alamos, NM 87545, USA}

\maketitle

\begin{abstract}
\begin{small}
 Scattering of the forward-shock synchrotron emission by a relativistic outflow located 
behind the leading blast-wave may produce an X-ray emission brighter than that coming 
directly from the forward-shock and may explain four features displayed by Swift X-ray 
afterglows: flares, plateaus (slow decays), chromatic light-curve breaks, and fast 
post-plateau decays. 
 For a cold scattering outflow, the reflected flux overshines the primary one if the 
scattering outflow is nearly baryon-free and highly relativistic. These two requirements 
can be relaxed if the scattering outflow is energized by weak internal shocks, so that 
the incident forward-shock photons are also inverse-Compton scattered, in addition to 
bulk-scattering. 
 Sweeping-up of the photons left behind by the forward shock naturally yields short X-ray 
flares. Owing to the boost in photon energy produced by bulk-scattering scattering, 
the reflected emission is more likely to overshine that coming directly from the forward 
shock at higher photon energies, yielding light-curve plateaus and breaks that appear only 
in the X-ray. The brightness, shape, and decay of the X-ray light-curve plateau depend 
on the radial distribution of the scatterer's Lorentz factor and mass-flux. 
 Chromatic X-ray light-curve breaks and sharp post-plateau decays cannot be accommodated 
by the direct forward-shock emission and argue in favour of the scattering-outflow model 
proposed here. On the other hand, the X-ray afterglows without plateaus, those with 
achromatic breaks, and those with very long-lived power-law decays are more naturally 
accommodated by the standard forward-shock model. Thus the diversity of X-ray light-curves 
arises from the interplay of the scattered and direct forward-shock emissions. 
\end{small}
\end{abstract}


\section{Introduction}

\begin{figure*}
\centerline{\psfig{figure=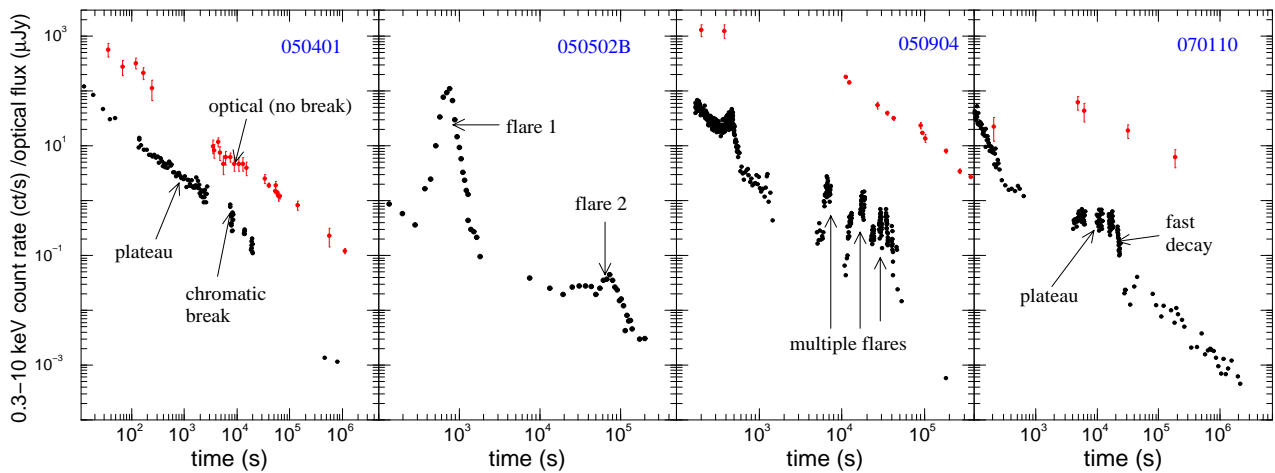,width=17cm}}
\caption{ Examples of Swift afterglows exhibiting the following X-ray features: 
    flares (GRB 050502B, 050904), plateaus (GRB 050401, 070110), fast post-plateau
    decay (GRB 070110), chromatic breaks (GRB 050401, 070110). Lower curves are
    the X-ray count rate (dots), upper curves are for the optical flux (red dots 
    with error bars).
}
\label{xlc}
\end{figure*}

 A large fraction of X-ray afterglows of Gamma-Ray Bursts (GRBs) monitored by Swift in 
the last 2.5 years exhibit (Figure \ref{xlc}) {\it flares} (short, brightening episodes 
during which the X-ray flux increases by a factor up to 1000) and {\it plateaus} 
(power-law flux decays slower than expected in the standard blast-wave model, for the 
measured slope of the X-ray spectrum). 

 The short duration $\delta t_{fwhm}$ of X-ray flares (Burrows et al 2007), 
half of which have $\delta t_{fwhm}/t \siml 0.1$ (Chincarini et al 2007),
is incompatible with an origin in the shock driven into the circumburst medium by 
the GRB ejecta (the {\it forward shock}) because its emission fluctuations cannot be
both very bright and short-lived. If the fluctuation's angular scale is $\delta \theta 
\simg \theta_{vis} \simeq \gamma^{-1}$, with $\gamma$ being the shock's Lorentz factor 
(i.e. larger than the area visible to the observer), then the flare should last for 
$\delta t \simg {\rm few}\times t$, which is the spread in the photon arrival-time 
caused by the curvature of the visible emitting surface. 
If the fluctuation's angular scale is less than $\theta_{vis}$  ($\delta \theta = 
\gamma^{-1}/n$, $n>1$), then the flare's duration, $\delta t \sim {\rm few}\times t/n$, 
can be arbitrarily small but the fluctuation's surface brightness would have to be 
$(1-10^3)n^2$ larger than that of the forward shock to overshine it and yield 
a flare that is 1--1000 times brighter than the underlying forward-shock flux. 

 The inability of the forward shock to produce short and bright flares has been taken as 
indication (e.g. Zhang et al 2006) that flares originate from the same mechanism as the 
burst: internal shocks in a fluctuating outflow (Rees \& M\'esz\'aros 1994).
This model requires that the central engine producing the GRB ejecta must live for 
a lab-frame duration comparable to the observer time when the flare is seen.
 
 The X-ray light-curve plateaus are generally smooth, thus a forward-shock interpretation
is plausible. That, during the plateau, the X-ray flux decays slower than expected in the 
standard forward-shock model (e.g. Paczy\'nski \& Rhoads 1993, M\'esz\'aros \& Rees 1997; 
Sari, Piran \& Narayan 1998) requires that at least one of the assumptions of this model 
is invalid. As the X-ray flux depends on the blast-wave energy and the electron \& magnetic 
field parameters in the forward shock, non-constancy of any of these quantities may explain 
the slow decay of X-ray plateaus: Nousek et al (2006), Panaitescu et al (2006a), and Zhang 
et al (2006) have proposed that energy is injected into the forward shock,
Ioka et al (2006) studied the light-curves arising from evolving microphysical parameters,
and Fan \& Piran (2006) analyzed the afterglow emission when all three quantities evolve.

 However, the existence of {\it chromatic} X-ray breaks at the end of some plateaus 
(e.g.  GRB afterglows 050401 and 070110 shown in Figure \ref{xlc}),
which are not exhibited by the optical light-curve, poses a serious problem to the above 
interpretation of plateaus because such a decoupling of the optical and X-ray emissions 
requires a strong spectral break to cross the X-ray range at the time of the chromatic 
break, yet the slope of the X-ray continuum is not observed to change across the break 
(Nousek et al 2006, Willingale et al 2007, Liang et al 2007). 

 The chromatic X-ray breaks may indicate that the afterglow emission arises from the 
reverse-shock crossing some inflowing, late ejecta that catch-up with the forward shock 
during the plateau. Again, to obtain the required decoupling of the optical and X-ray 
light-curves, a spectral break must be in between those domain: 
Uhm \& Beloborodov (2007) attributed that break to the cooling of reverse-shock electrons, 
while Genet, Daigne \& Mochkovitch (2007) identified it with the characteristic synchrotron 
frequency at which electrons would radiate if only 1 percent of reverse-shock electrons 
acquired equipartition energies.
It remains to be investigated if, by tracking the cooling of forward-shock electrons during
energy injection or by assuming that only 1 percent of them reach equipartition, the 
forward-shock acquires the same ability to decouple the optical and X-ray light-curves 
as the reverse shock may have. 

 In this paper, we continue to focus on the forward-shock emission, which naturally explains 
the power-law decay of afterglow light-curves, and investigate the possibility that the 
decoupling of the optical and X-ray emissions is due to a substantial contribution
to the X-ray flux from an incoming, delayed outflow which up-scatters the forward-shock 
emission. Half of the synchrotron photons emitted by the forward shock are left behind it 
(the half which, in the shock frame, are emitted at an angle $\alpha' > \pi/2$ relative
to the radial direction of motion) and boosted relativistically by a factor ranging from 
$\gamma$ (for $\alpha'=\pi/2$) to $(2\gamma)^{-1}$ (for $\alpha'=\pi$). The former are 
caught-up with by the incoming outflow if its Lorentz factor $\Gamma$ is larger than that
of the forward-shock; the latter will always be reached by the scattering outflow; for 
either, the Doppler boost after scattering is a factor $(\Gamma/\gamma)^2$ higher than 
that of a forward-shock photon travelling directly toward the observer (i.e. the primary 
emission)

 There are two reasons for which the up-scattered emission may be brighter than the 
primary. One is that the photon travel from place of emission to that of scattering 
delays the scattered emission relative to the direct forward-shock emission.
Given that the latter decays, it follows that the scattered flux arriving at some time 
may be brighter than the forward-shock flux at same time. 
 The second reason is at work if $\Gamma > \gamma$. In this case, scattering increases
the photon energy more than the Doppler boost of the forward-shock does, so that, 
if the energy of the seed photon is above the peak of the synchrotron spectrum of 
the forward-shock emission, then the scattered flux at some observer-frame photon energy 
may exceed that of the forward-shock at same photon energy.

 There is also one reason for which the scattered emission could be dimmer than
the forward-shock's: for plausible kinetic energies of the scattering outflow (below
about $10^{53}$ erg) and the range of radii over which the up-scattering takes place 
(the forward-shock radius is $10^{16}-10^{17.5}$ cm), the scattering outflow is optically 
thin, thus only a small fraction of the photons left behind by the forward shock will be 
swept-up. 

 We assume that the scattering electrons are cold (i.e., in the frame of the incoming 
outflow, the primary photons do not gain energy through electron scattering). 
In this case, for the optical depth to electron scattering to be sufficiently large to 
yield a scattered flux brighter (at higher photon energies) than the forward-shock's,
the scattering fluid must consist mostly of electron-positron pairs (i.e. baryon-poor), 
to maximize the number of leptons for a given kinetic energy and Lorentz factor of the 
scattering outflow. 
 
 Such a pair-enriched outflow are not expected from radiation-dominated fireballs, owing 
to pair-annihilation above the photospheric radius, but could result in electromagnetic 
(Poynting) outflows (Lyutikov 2006) from the dissipation of magnetic around the 
deceleration radius. The latter model has received support from the timing of fast-decaying 
GRB tails and the epoch when the forward-shock emission emerges, which suggest that 
the burst and early afterglow emissions are produced at comparable radii ($10^{15-16}$
cm, Kumar et al 2007). This is a plausible feature of the electromagnetic model and 
disfavours the internal-shock model for GRBs for which the subsecond burst variability
timescale requires the GRB emission to be produced at $10^{13-14}$ cm. Alternatively,
if the short burst variability can be produced at larger radii, then the proximity of 
the locations where the burst and early afterglow emissions are released would point
to internal-external shocks occurring between some incoming ejecta and the decelerating
forward-shock (Fenimore \& Ramirez-Ruiz 1999; Ramirez-Ruiz, Merloni \& Rees 2001).

 However, if the scattering electrons are hot (e.g. accelerated by internal shocks), 
then inverse-Compton scattering of forward-shock photons will boost into the X-rays 
seed photons of even lower energy, which may be numerous enough to compensate for the 
lower optical thickness of a scattering outflow with a normal baryon-to-lepton load. 
Thus, a hot outflow with a normal proton-to-electron composition (as expected for 
fireballs accelerated by radiation pressure) may still yield a scattered flux overshining 
in the X-rays that from the forward shock.

\section{Set-up}
\label{setup}

 {\it Notations}: primed quantities are measured in the frame of the emitting fluid,
unprimed are in the lab or observer frames; subscripts and superscripts "$sc$" and "$fs$" 
refer to scattering fluid and forward-shock quantities, respectively; $\Gamma$ and $R$ 
are the scatterer's Lorentz factor and radius, $\gamma$ and $r$ are those of the forward 
shock; $t$ denotes lab-frame time, $T$ is for observer (photon arrival) time.  

 A scattering, relativistic outflow with a significant radial spread can be approximated 
as a sequence of shells of zero radial thickness, thus we begin by calculating the
emission from a scattering surface. Both the forward shock and scattered emissions are
integrated over the equal arrival-time surface defined by $T = t - r \cos \omega$ 
where $\omega$ is the angle between the direction of motion of the radiating fluid element 
and the center of explosion -- observer axis. For analytical calculations, one can ignore
the spread in the photon arrival-time caused by the spherical curvature of the emitting 
surface and restrict attention to $\omega = 0$ (i.e. assume that the fluid is moving
directly toward the observer). The only important effect of the surface curvature which
must be taken into account that is that, for a source of Lorentz factor $g$, the 
observer receives emission from a patch of angular opening $\omega_{max} = g^{-1}$ 
centered on the center--observer axis, as the emission from angles $\omega > g^{-1}$ 
is less boosted relativistically. This means that the specific flux is 
\beq
 F(\nu) \propto \omega_{max}^2 g^3 I'(\nu/g) \propto g I'(\nu/g)
\label{beam}
\eeq
where $\omega_{max}^2$ gives the source solid angle and $g^3 I'(\nu/g)$ is the intensity 
of the relativistically-beamed emission of comoving frame intensity $I'$.   
  
\begin{figure}
\centerline{\psfig{figure=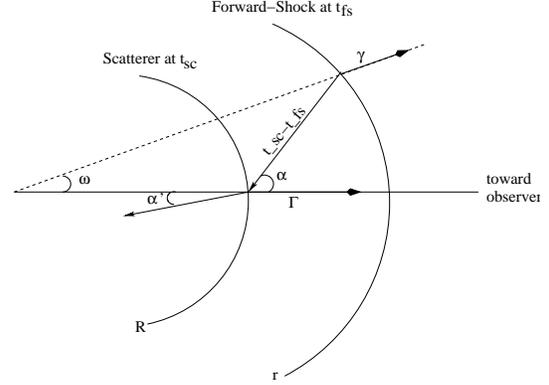,width=7cm}}
\caption{Geometry of scatterer and forward shock, moving at Lorentz factors $\Gamma$
    and $\gamma$, respectively. A forward-shock photon emitted at lab-frame time 
    $t_{fs}$ and moving at lab-frame angle $\alpha$ (relative to the radially-out
    direction of flow) enters the scattering shell at time $t_{sc}$ and scatterer-frame 
    angle $\alpha'$.  }
\label{geom}
\end{figure}

 In the frame of the scattering outflow, the intensity $I'_{sc}$ of the radiation scattered 
in the radial direction is the integral over all incident angles $\alpha'$ (see Figure \ref{geom})
of the incoming radiation intensity $I'_{in}$ times the electron cross-section for a photon 
deflection-angle $\pi - \alpha'$ (i.e. the dipole pattern of electron scattering):
\beq
 I'_{sc} = \frac{3}{8} \tau_{sc} \int_0^\pi d\alpha' \sin \alpha' (1+\cos^2\alpha') 
           I'_{in}(\alpha')
\label{Isc}
\eeq   
where $\tau_{sc} < 1$ is the optical thickness to electron-scattering of the scattering 
surface. For a scatterer of kinetic energy $E_{sc}=10^{53}E_{sc,53}$, Lorentz factor 
$\Gamma=10^3\Gamma_3$, radius $R=ct=3 \times 10^{16}t_6$ cm, and $N_e/N_p$ leptons per 
baryon
\beq
 \tau_{sc} = 0.72 \times 10^{-2} \frac{E_{sc,53}}{\Gamma_3 t_6^2} 
          \left( 1 + \frac{N_p m_p}{N_e m_e} \right)^{-1}    \;.
\label{tau}
\eeq
 If the curvature of the emitting surface is taken into account (as done numerically), 
the term $1+\cos^2\alpha'$ of equation (\ref{Isc}) is replaced by $1+\cos^2\alpha'\cos^2\omega'
+\frac{1}{2}\sin^2\alpha'\sin^2\omega'$ where $\omega'$ is the comoving-frame angle 
corresponding to the lab-frame $\omega$: $\cos \omega' = (\cos\omega-B)/(1-B\cos\omega)$, 
with $B=(1-\Gamma^{-2})^{-1/2}$ being the {\it constant} velocity of the scattering surface.

 The intensity $I'_{in} (\alpha')$ of the incident radiation is the comoving intensity 
$I'_{fs}$ of the forward-shock emission beamed relativistically by the relative motion
of the scatterer and forward-shock. Considering specific intensities,
\beq
 I'_{in}(\nu',\alpha') = D^3(\alpha) I'_{fs}(\nu'/D,\alpha) 
\label{Iin}
\eeq
where 
\beq
 D (\alpha) = \frac {\Gamma(1+B\cos\alpha)} {\gamma[1+b\cos(\alpha-\omega)]}
\label{D}
\eeq
is the Doppler boost of the scatterer--forward-shock relative motion for a forward-shock
photon moving at a lab-frame angle $\pi - \alpha$ relative to the scatterer direction of
motion, thus
\beq
 \cos \alpha = \frac{\cos\alpha'-B}{1-B\cos\alpha'} \;. 
\eeq
$b$ is the forward-shock velocity at the retarded time $t_{fs}(t_{sc},\alpha)$ when 
the photon arriving at scatterer at time $t_{sc}$ and angle $\alpha$ was emitted, 
and $\omega(t_{fs},\alpha)$ is the angle between the radial direction of motion of the
forward-shock patch at angle $\alpha$ and time $t_{fs}$ and the direction toward the 
observer.

 The kinematics of the photon motion from its place of emission ($r,t_{fs}$) to that of
scattering ($R,t_{sc}$) provides two equations 
\beq
 r(t_{fs}) \cos\omega = R(t_{sc}) + (t_{sc}-t_{fs}) \cos\alpha 
\eeq
\beq
 r(t_{fs}) \sin\omega = (t_{sc}-t_{fs}) \sin\alpha 
\eeq
from where the motion angle $\omega$ and radius $r(t_{fs})$ at the retarded time $t_{fs}$ 
can be calculated. For the latter:
\beq
 r^2 = R^2 + 2R(t_{sc}-t_{fs})\cos\alpha + (t_{sc}-t_{fs})^2
\label{rr}
\eeq
can be solved for $t_{fs}$ using the equations for the scatterer's kinematics:
\beq
 R(t_{sc}) = B (t_{sc}-t_{lag}) \simeq \left( 1-\frac{1}{2\Gamma^2} \right) (t_{sc}-t_{lag}) 
\eeq
where $t_{lag}$ is the time elapsed between the ejection of the leading outflow producing
the forward shock and the lagging scattering outflow, and for the forward-shock's motion:
\beq
 r(t_{fs}) = \int_0^{t_{fs}} b(t)dt \simeq t_{fs} - \int_0^{t_{fs}} \frac{dt}{2\gamma^2(t)}\;.  
\label{rfs}
\eeq
The forward-shock dynamics resulting from sweeping-up and energizing a circumburst medium
of particle density
\beq
 n(r) \propto r^{-s} 
\eeq
is $\gamma = \gamma_0$ until a "deceleration time" $t_0$ when the reverse-shock crosses 
the relativistic ejecta, followed by
\beq
 \gamma(t) = \gamma_0 \left( \frac{t}{t_0} \right)^{-(3-s)/2} \quad  (s < 3) \;.
\label{gm}
\eeq
Then, equation (\ref{rfs}) yields 
\beq
 r(t_{fs}) \simeq t_{fs} \left[ 1 - \frac{1}{2(4-s)\gamma^2(t_{fs})} \right] \;.
\label{rFS}
\eeq
Substituting equations (\ref{gm}) and (\ref{rFS}) in (\ref{rr}) leads to the following
equation for $t_{fs}(t_{sc},\alpha)$ :
\beq
 2 t_{fs} \left[ R\cos\alpha + t_{sc} - T_0 \left(\frac{t_{fs}}{t_0}\right)^{4-s} \right] = 
      R^2 + 2Rt_{sc}\cos\alpha +t_{sc}^2
\label{tfs}
\eeq
where $T_0 \equiv t_0/[2(4-s)\gamma_0^2]$ is the observer-frame deceleration time (i.e.
the arrival-time of photons emitted by the forward-shock at the onset of its deceleration).

 From equation (\ref{beam}), the flux received by the observer at time $T_{sc}$ satisfies
\beq
 F_{sc}(\nu,T_{sc}) \propto \Gamma I'_{sc}(\nu/\Gamma,t_{sc}) 
\label{Fsc}
\eeq
where 
\beq
 T_{sc} = t_{sc} - R(t_{sc}) = t_{lag} + \frac{1}{2\Gamma^2} (t_{sc}-t_{lag}) \;.
\label{Tscat}
\eeq
The flux received directly from the forward shock satisfies
\beq
 F_{fs}(\nu,T_{fs}) \propto \gamma(t_{fs}) I'_{fs}(\nu/\gamma,t_{fs}) 
\label{Ffs}
\eeq
with
\beq
 T_{fs} = t_{fs} - r(t_{fs}) \simeq \frac{1}{2(4-s)} \frac{t_{fs}}{\gamma^2(t_{fs})} 
\eeq
which, together with equation (\ref{gm}), leads to
\beq
 T_{fs} = T_0 \left( \frac{t_{fs}}{t_0} \right)^{4-s} , \quad 
 \gamma (T_{fs}) = \gamma_0 \left( \frac{T_{fs}}{T_0} \right)^{-(3-s)/(8-2s)} \;.
\label{gfs}
\eeq
It can be shown that, for times after the deceleration timescale $T_0$, the forward-shock 
Lorentz factor is
\beq
 \gamma (T_{fs}) \simeq 450 \left( \frac{E_{fs,53}}{n_0} \right)^{1/8} T_{fs}^{-3/8}
\label{g0}
\eeq
for a homogeneous medium ($s=0$) and
\beq
 \gamma (T_{fs}) \simeq 135 \left( \frac{E_{fs,53}}{A_*} \right)^{1/4} T_{fs}^{-1/4}
\label{g2}
\eeq
for a wind-like medium ($s=2$),
with $T_{fs}$ is measured in seconds, $E_{fs,53}$ the forward-shock kinetic energy
in $10^{53}$ erg, $n_0$ the circumburst medium density in protons/$\rm{cm^3}$, 
$A_* = 1$ for the wind blown by a GRB progenitor with a mass-loss rate of $dM_w/dt
=10^{-5} M_\odot \,\rm{yr^{-1}}$ and a terminal wind velocity of $v = 1000\; 
{\rm km\;s^{-1}}$, and $A_* \propto (dM_w/dt)/v$.

 Equations (\ref{Isc}), (\ref{Iin}), (\ref{Fsc}), and (\ref{Ffs}) relate the scattered
flux with that received directly from the forward shock. To complete their numerical 
calculation, the comoving frame intensity $I'_{fs}(\nu')$ of the forward-shock emission 
must be specified. For optically thin synchrotron, that intensity is
\beq
 I'_{fs} (\nu') = I'_{fs} (\nu'_p) \left\{ \begin{array}{ll} 
      \hspace*{-2mm} (\nu'/\nu'_i)^{1/3} & \nu' < \nu'_i \\
      \hspace*{-2mm} (\nu'_i/\nu')^{(p-1)/2} & \nu'_i < \nu' < \nu'_c \\
      \hspace*{-2mm} (\nu'_c/\nu')^{1/2} & \nu'_c < \nu' < \nu'_i \\
      \hspace*{-2mm} (\nu'_i/\nu'_c)^{(p-1)/2} (\nu'_c/\nu)^{p/2} & \nu'_i, \nu'_c < \nu' 
   \end{array} \right.
\label{spek}
\eeq
where $p$ is the exponent of the power-law distribution of electrons with energy 
$\epsilon$ in the forward shock -- $dn/d\epsilon \propto \epsilon^{-p}$ -- and
$\nu'_i$ is the "injection" frequency (the characteristic synchrotron frequency at 
which the electrons of minimal energy radiate),
$\nu'_c$ is the "cooling" frequency (the characteristic frequency at which radiate 
the electrons whose radiative timescale equals the dynamical one), and
$I'_{fs}(\nu'_p)$ is the specific intensity at the peak $\nu'_p = \min(\nu'_i,\nu'_c)$ 
of the forward-shock emission spectrum. 
Assuming that electrons and magnetic fields acquire constant fractions of the internal
energy of the shocked medium, it can be shown that the spectral characteristics of
the forward-shock emission have the following evolutions:
\beq
 (t_{fs} < t_0) \quad I'_{fs} (\nu'_p) \propto t_{fs}^{3-1.5s} , 
    \;\; \nu'_i \propto t_{fs}^{-0.5s} ,\;\; \nu'_c \propto t_{fs}^{1.5s-2} 
\label{Ipk0}
\eeq
before deceleration and
\beq
 (t_{fs} > t_0) \quad I'_{fs} (\nu'_p) \propto t_{fs}^{1.5-s} ,
     \;\;  \nu'_i \propto t_{fs}^{s-4.5} ,\;\; \nu'_c \propto t_{fs}^{s-0.5} 
\label{Ipk1}
\eeq
after deceleration.
In general, scattering of the pre-deceleration forward-shock emission yields a 
negligible contribution to the total scattered flux because, before deceleration,
the forward-shock bolometric emission increases with time, thus the incident flux is 
largest at deceleration or after that.

\section{X-ray flares from scattering surfaces}
\label{surface}

 With the exception of $\alpha' \simeq 0$ photons, the last term in the square bracket 
of equation (\ref{tfs}) is comparable to the sum of the first two and cannot be ignored.
This prevents us to solve analytically for $t_{fs}(t_{sc},\alpha')$, hence the analytical
integration of the scattered emission is not possible either. 
 Nevertheless, some properties of the scattered emission decay can be obtained by 
considering only the $\alpha'= 0$ photons, i.e.  photons which are emitted by the forward 
shock at angle $\pi$ relative to its direction of motion and then scattered along the 
center--observer axis. For these photons, the last term in the square bracket of equation 
(\ref{tfs}) can be ignored, leading to 
\beq
 t_{fs}(\alpha'=0) = t_{sc} - \frac{1}{2} t_{lag} \;.
\label{tfs0}
\eeq 

 Because the scattered pre-deceleration emission is negligible, we can restrict attention
to $t_{fs} > t_0$. Then, given that $t_{sc} > t_{fs}$, it follows that $t_0 \gg t_{lag}$ 
ensures that $t_{sc} \gg t_{lag}$. For a Newtonian reverse shock, the lab-frame deceleration 
timescale corresponds to the mass swept-up by the forward shock reaching a fraction 
$\gamma_0^{-1}$ of the ejecta, which leads to
\beq
 (s=0) \quad t_0 = 4.3\times 10^6 (E_{sc,53}/n_0)^{1/3} (\gamma_0/100)^{-2/3} \; {\rm s}
\eeq
\beq
 (s=2) \quad t_0 = 1.3\times 10^5 (E_{sc,53}/A_*) (\gamma_0/100)^{-2} \; {\rm s}
\eeq
hence the $t_0 \gg t_{lag}$ condition sets an upper on the forward-shock's initial Lorentz
factor:
\beq
 (s=0) \quad \gamma_0 < 10^6 (E_{sc,53}/n_0)^{1/2} (t_{lag}/10^4\,{\rm s})^{-3/2}
\label{glim0}
\eeq
\beq
 (s=2) \quad \gamma_0 < 360\, (E_{sc,53}/A_*)^{1/2} (t_{lag}/10^4\,{\rm s})^{-1/2} \;.
\label{glim2}
\eeq
Anticipating the result that the scattered emission arrives at observer time $T_{sc}
\simeq t_{lag}$, we have scaled $t_{lag}$ to the latest time when X-ray flares
are usually observed by the X-ray Telescope (XRT) on Swift. 
Thus, for $\gamma_0$ satisfying condition (\ref{glim0})
or (\ref{glim2}), $t_{sc} \gg t_{lag}$ is satisfied and $t_{lag}$ can be ignored 
in the round bracket of equation (\ref{Tscat}) and in equation (\ref{tfs0}), leading to 
\beq
 t_{fs}(\alpha'= 0) \simeq t_{sc} 
\label{tfstsc}
\eeq
and
\beq
  T_{sc} \simeq t_{lag}+\frac{t_{sc}}{2\Gamma^2} \simeq t_{lag}+\frac{t_{fs}}{2\Gamma^2} 
  = t_{lag} + (4-s) \frac{\gamma^2(T_{fs})}{\Gamma^2} T_{fs} 
\label{Tsc}
\eeq
where the last result follows from equation (\ref{gfs}). Equation (\ref{Tsc}) relates the
arrival-time of a forward-shock photon to when it would arrive if it were emitted backwards
($\alpha = 0$ corresponds to angle $\pi$ between the photon propagation direction and that 
of the forward shock) and then scattered. It allows the calculation of the range of
arrival times for scattered photons, starting with the photons emitted by the forward
shock at the onset of deceleration ($T_0$)
\beq
 T_{sc}^{min} = t_{lag} + (4-s) \frac{\gamma_0^2}{\Gamma^2} T_0
\label{Tmin}
\eeq
and until the scatterer catches-up with the forward shock. The latter is given by
setting $T_{fs} = T_{sc}$ in equation (\ref{Tsc}) which leads to
\beq
 T_{sc}^{max} = t_{lag} + (4-s) \frac{\gamma^2(T_{sc})}{\Gamma^2} t_{lag} 
\label{Tmax}
\eeq
where $\gamma(T_{sc}) \ll \Gamma$ was assumed, which is ensured if $\gamma_0 \ll \Gamma$.
The latter will always be the case considered here because only then the scattered emission 
can be brighter than that from the forward shock. Thus, for a forward-shock initial Lorentz 
factor well below that of the scattering surface, equations (\ref{Tmin}) and (\ref{Tmax}) 
show that the forward-shock photons left behind the shock and swept-up by the scatterer
arrive at observer at $T \simeq t_{lag}$ over a time interval
$\delta T \simeq (4-s) [\gamma^2(T)/\Gamma^2] t_{lag}$ which is much less than $t_{lag}$. 

  This means that the scattering of the forward shock emission by a surface of higher 
Lorentz factor can yield a flare of duration
\beq
 \delta T \simeq  (4-s) \frac{\gamma^2(T)}{\Gamma^2} T \ll T
\label{delT}
\eeq
{\it much shorter} than the age of the afterglow.
 Furthermore, that the flare from a scattering surface occurs at $T \simeq t_{lag}$, 
shows that, in the case of a radially-extended scattering outflow, the brightness of 
the scattered emission at observer time $T$ reflects the properties (Lorentz factor 
$\Gamma$, mass distribution $dM_{sc}/dt_{lag}$) of the scattered fluid ejected at lab-frame 
time $t_{lag} = T$, i.e. {\it the brightness of the scattered emission mirrors the 
scattering outflow properties in real time} (modulo cosmological time-dilation).
 
 For the calculation of the scattered emission, we approximate $\omega \simeq 0$ in 
equation (\ref{D}), hence the Doppler factor for the forward-shock--scatterer relative 
motion is 
\beq
 D = \Gamma/\gamma \;.
\label{DD}
\eeq
This means that, in the frame of the scatterer, the forward-shock 
photons arrive within an angle $D^{-1} = \gamma/\Gamma$ around the radial direction of 
motion. Scattering by electrons of the incoming photons redistribute them nearly 
isotropically, hence the comoving frame of the scattered radiation (equation \ref{Isc}) 
satisfies
\beq
 I'_{sc} \simeq \tau_{sc} \frac{\pi(\gamma/\Gamma)^2}{4\pi} I'_{in} \;.
\eeq
 Substituting in equation (\ref{Fsc}), we obtain
\beq
 F_{sc}(\nu,T_{sc}) \propto \Gamma \tau_{sc} \left(\frac{\gamma}{\Gamma}\right)^2 
     I'_{in}(\nu/\Gamma,t_{sc}) 
\eeq
which, after using equations (\ref{Iin}) and (\ref{DD}), yields 
\beq 
 F_{sc}(\nu,T_{sc}) \propto  \tau_{sc} \frac{\Gamma^2}{\gamma} 
   I'_{fs}(\nu\gamma/\Gamma^2,t_{sc}) \;.
\eeq
Then, with the aid of equation (\ref{Ffs}), we can write
\beq
 \frac{F_{sc}(\nu,T_{sc})}{F_{fs}[\nu,T_{fs}(T_{sc})]} \simeq 
        \tau_{sc}(T_{sc}) \frac{\Gamma^2}{\gamma^2(T_{fs})}
       \frac {I'_{fs}[\nu\gamma/\Gamma^2,t(T_{sc})]} {I'_{fs}[\nu/\gamma,t(T_{sc})]} 
\label{frac1}
\eeq
where we indicated that the forward-shock emission released at lab-frame time $t(T_{sc})$ 
arrives at $T_{sc}$ after being scattered and at time $T_{fs}(T_{sc})$ if it is not scattered 
(the direct emission). 
In the simpler case when the forward-shock emission has a power-law spectrum between the 
two frequencies in the $rhs$ of equation (\ref{frac1}) (i.e. there are no spectral breaks
in between), $I'_{fs} (\nu) \propto (\nu')^{-\beta}$, the ratio of intensities in that 
equation is $(\Gamma/\gamma)^{2\beta}$, thus
\beq
 \frac{F_{sc}(\nu,T_{sc})}{F_{fs}[\nu,T_{fs}(T_{sc})]} \simeq
        \tau_{sc} \left( \frac{\Gamma}{\gamma} \right)^{2\beta+2} \;.
\label{frac2}
\eeq

 This result can also be derived from that the ratio of the spectral peak fluxes and peak 
frequencies of the flare and forward-shock emissions 
are
\beq
 \frac{F_{sc}(\nu_p^{sc})}{F_{fs}(\nu_p^{fs})} = \tau_{sc}  \frac{\Gamma^2}{\gamma^2}\; ,
  \quad  \frac{\nu_p^{sc}}{\nu_p^{fs}} = \frac{\Gamma^2}{\gamma^2}
\label{boost}
\eeq
and from that the forward-shock and scattered fluxes at a frequency $\nu > \nu_p$ are
given by $F(\nu) = F_p (\nu/\nu_p)^{-\beta}$.
For either ratio of equation (\ref{boost}), one factor $\Gamma/\gamma$ results from the 
relativistic motion of the sources toward the observer and another one is from the 
forward-shock--scatterer relative motion. For the peak flux ratio, the factor 
$(\Gamma/\gamma)^2$ represents the time-contraction (the angular beaming part associated 
with the relative motion is lost because electron scattering makes the incoming radiation
nearly isotropic, while the frequency Doppler boosts are lost because we are working with fluxes 
per unit frequency). For the peak frequency ratio, the factor $(\Gamma/\gamma)^2$ represents 
the relativistic Doppler effect.   

\begin{figure}
\psfig{figure=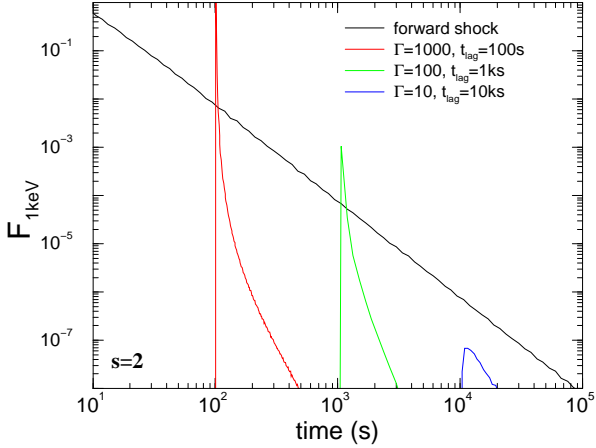,width=8cm}
\caption{Forward-shock and scattered X-ray light-curves (in arbitrary units)
   from a scattering surface of Lorentz factor $\Gamma$, ejected at $t_{lag}$ after 
   the forward shock. The optical thickness to electron scattering of the reflecting
   surface is calculated from equation (\ref{tau}) for $E_{sc}=10^{53}$ erg and
   a purely leptonic composition ($N_p=0$). The Lorentz factor $\gamma$ of the 
   forward-shock is that given in equation (\ref{g2}) for $E_{fs}/A_*=10^{53}$ erg.
   The evolution of the forward-shock spectral properties are those of equations 
   (\ref{Ipk0}) and (\ref{Ipk1}) for $s=2$. The comoving-frame spectrum of the 
   forward-shock emission is that given in the upper two lines of equation (\ref{spek}) 
   with $p=3$, observer-frame injection frequency $h\nu_i (1s) = 100$ eV, and cooling 
   frequency well above X-rays.
   Because $\Gamma \gg \gamma$ for the earlier two flares, these flares
   (1) occur at an observer time equal to the lab-frame delay time between the 
      forward shock and scatterer ejections, 
   (2) last much shorter than the time when they occur, and
   (3) are brighter than the forward-shock emission.  }
\label{flares}
\end{figure}

 Figure \ref{flares} shows the flares obtained numerically for a few values of the 
scattering surface Lorentz factors $\Gamma$ and for a wind-like medium. It illustrates
that the flare occurs at an observer time equal to the initial lag of the scattering 
surface behind the forward-shock (equations \ref{Tmin} and \ref{Tmax}), lasting
much shorter than the time when it occurs (equation \ref{delT}), and being brighter
than the underlying forward-shock emission for a larger $\Gamma/\gamma$ Lorentz factor
contrast (equation \ref{frac2}). This condition implies that the emission scattered 
along the observer--center line ($\alpha=0$) arrives at observer over a very short 
time (equation \ref{delT}), hence the decay of the flare from a fast scattering
surface is set by the geometrical curvature of the surface (photons emitted from 
fluid moving at increasing angles relative to observer arrive progressively later), 
which leads to a sharp 
rise and a $F_{sc}(\nu) \propto (T_{sc}-t_{lag})^{-(\beta+2)}$ flare decay (Kumar \& 
Panaitescu 2000). Other kinds of flare profiles can be obtained with a scattering 
outflow of radial extent $\delta t_{lag} \simg t_{lag}$ and adequate distributions 
for the scatterer Lorentz factor $\Gamma(t_{lag})$ and mass $dM_{sc}/dt_{lag}$ 
(\S\ref{shapes}).

\section{X-ray plateaus from scattering outflows -- analytical treatment}
\label{analytical}

 To calculate from equation (\ref{frac1}) the emission scattered by a radially-extended 
outflow with a $dM/dt_{lag}$ radial mass-distribution, we discretize it as a sequence 
of scattering surfaces of mass $\delta M = \delta t_{lag}(dM/dt_{lag})$ ejected at an 
interval $\delta t_{lag}$ and add their scattered emissions. Thus, the optical thickness 
of the discretized outflow is
\beq
 \tau_{sc}(T_{sc}) \propto \frac{\delta t_{lag} (dM/dt_{lag})}{R^2(T_{sc})} 
\label{tausc}
\eeq
where 
\beq
  R(T_{sc}) \simeq t_{sc} - t_{lag} \simeq 2\Gamma^2 (T_{sc}-t_{lag})  
\label{Rsc}
\eeq
as can be inferred from equation (\ref{Tscat}) in the $\Gamma \gg 1$ limit.
 The power-law spectrum of the forward-shock emission (equation \ref{spek}) and the
power-law evolution of its spectral characteristics (equation \ref{Ipk1}), lead to a 
received forward-shock emission that decays as a power-law in time
\beq
 F_{fs}(\nu) \propto T^{-\alpha_{fs}}
\label{alfafs}
\eeq
where the index $\alpha_{fs}$ is a linear function of the spectral slope $\beta$ that
will be specified later.
By substituting equations (\ref{tausc}), (\ref{Rsc}), and (\ref{alfafs}) in (\ref{frac1})
we arrive at
\beq
 F_{sc}(\nu,T_{sc}) \propto \frac{\delta t_{lag} (dM/dt_{lag})}{\Gamma^2\gamma^2 (T_{sc}-t_{lag})^2}
         \frac{I'_{fs}(\nu\gamma/\Gamma^2)} {I'_{fs}(\nu/\gamma)} T_{fs}^{-\alpha_{fs}} \;.
\label{fsc}
\eeq

 From the first equation (\ref{gfs}) and equation (\ref{tfstsc}), and using $t_{sc}$ 
obtained from equation (\ref{Tscat}) in the $t_{sc} \gg t_{lag}$ limit (which is a 
good approximation), the arrival-time $T_{fs}$ of the direct forward-shock emission
corresponding to an arrival-time $T_{sc}$ of the scattered photons is found to satisfy
\beq
  T_{fs} \propto [\Gamma^2 (T_{sc}-t_{lag})]^{4-s} \;.
\label{Tfs}
\eeq

 For the scattered flux at time $T$, we integrate the flux from the scattering outflow 
element $\delta M$ to obtain the fluence $\Phi_{sc}(\nu,T)$ and divide it by the observer 
time-interval $\delta T$ over which that fluence is spread. Taking into account the 
one-to-one correspondence between the lag time $t_{lag}$ and the observer arrival-time 
$T$ of scattered emission, we obtain that, for an extended scattering outflow, 
\begin{displaymath}
 F_\nu^{sc}(T) = \frac{1}{\delta t_{lag}} \int\limits_{T_{sc}^{min}(T)}^{T_{sc}^{max}(T)}
              \hspace*{-4mm} F_{sc}(\nu,T_{sc}) dT_{sc} \propto 
              \frac{dM}{dT} \Gamma^{4-2s-(8-2s)\alpha_{fs}} 
\end{displaymath}
\beq
  \quad  \quad \quad  \times 
   \int\limits_{T_{sc}^{min}(T)}^{T_{sc}^{max}(T)} 
          \hspace*{-4mm} (T_{sc}-t_{lag})^{1-s-(4-s)\alpha_{fs}}
        \frac {I'_{fs}(\nu\gamma/\Gamma^2)} {I'_{fs}(\nu/\gamma)}\; dT_{sc} 
\label{Fnu}
\eeq   
after using equations (\ref{fsc}) and (\ref{Tfs}), with the forward-shock Lorentz factor
from the second equation (\ref{gfs}).

 To continue, we have to specify the location of the two break frequencies 
(injection $\nu'_i$ and cooling $\nu'_c$) of the forward-shock synchrotron spectrum.  
The X-ray light-curve plateaus occur after about 1 ks, when the injection frequency 
is expected to be below X-rays ($\nu'_i < \nu_x/\gamma$) even for equipartition 
electron and magnetic field parameters (for several afterglows, the optical spectral 
energy distribution at $T < 1$ ks decreases with photon energy, thus $\nu_i$ is even
lower, below optical).
 The up-scattered injection frequency could be below or above X-rays. 
In the latter situation ($\nu'_i > \nu_x\gamma/\Gamma^2$), we are interested only in 
the case where the up-scattered cooling frequency is below X-rays ($\nu'_c < 
\nu_x\gamma/\Gamma^2$) because, in the opposite case, the up-scattered spectrum would 
be $F_\nu \propto \nu^{1/3}$, which has never been observed (afterglow spectra measured 
by XRT are softer than $\nu^{-1/2}$).

\subsection{ $\nu'_i < \nu_x\gamma/\Gamma^2 < \nu_x/\gamma < \nu'_c$ }
\label{case1}

 In this case, $\nu_i^{fs} < \nu_x < \nu_c^{fs}$ and $\nu_i^{sc} < \nu_x < 
\nu_c^{sc}$, thus the spectral slopes of the forward-shock and scattered X-ray emissions 
are $\beta_x^{fs}=\beta_x^{sc}=(p-1)/2$, with $p$ the exponent of the electron 
distribution with energy in the forward shock. The decay index of the forward-shock
X-ray emission is
\beq
 \alpha_{fs} = \frac{3}{2} \beta_x + \frac{s}{8-2s} 
\eeq 
where $\beta_x \equiv (p-1)/2$.

 The ratio of intensities in equation (\ref{Fnu}) is $(\Gamma/\gamma)^{2\beta_x}$, 
which, after using equations (\ref{Tmin}) and (\ref{Tmax}) for the integral limits, 
leads to
\beq
 F_\nu^{sc}(T) \propto \frac{dM}{dT} \Gamma^{2\beta_x} \frac{1}{A} 
              \left\{ [T\gamma^2(T)]^A - (T_0\gamma^2_0)^A \right\}
\label{int}
\eeq
where
\beq
  A = 2 - 1.5s - \beta_x (3-0.5s) \;.
\eeq
For $\beta_x < \beta_0$ defined by
\beq
 \beta_0 = \frac{4-3s}{6-s} = \left\{ \begin{array}{ll} 
          2/3 & s=0 \\ -1/2 & s=2 \end{array}  \right. 
\label{b01}
\eeq
the exponent $A$ is positive and the scattered flux of equation (\ref{int}) is 
dominated by the scattering of forward-shock emission at $T_{sc} \siml T$. 
Taking into account that $T\gamma^2(T) \propto T^{1/(4-s)}$, it follows that
\begin{displaymath}
 F_\nu^{sc}(T) \propto \frac{dM}{dT} \Gamma^{2\beta_x} T^{A/(4-s)} 
\end{displaymath}
\beq
    \quad \quad \quad \quad =
           \frac{dM}{dT}  \Gamma^{2\beta_x} \left\{ \begin{array}{ll} 
               \hspace*{-2mm} T^{-(3\beta_x-2)/4} & \hspace*{-2mm} s=0 \\ 
               \hspace*{-2mm} T^{-(\beta_x+1/2)} & \hspace*{-2mm} s=2 
          \end{array} \right. \;.
\label{F1}
\eeq
For $\beta_x > \beta_0$, $A$ is negative and the scattered flux is dominated by the 
scattering of photons produced by the forward shock around the deceleration time.
In this case, (\ref{int}) leads to
\beq
  F_\nu^{sc}(T) \propto \frac{dM}{dT} \Gamma^{2\beta_x}
\label{F0}
\eeq
which is, evidently, independent of the ensuing decay of the forward-shock emission.

\subsection{ $\nu'_i < \nu_x\gamma/\Gamma^2 < \nu'_c < \nu_x/\gamma$ }

 This case corresponds to $\nu_i^{fs} < \nu_c^{fs} < \nu_x$ and $\nu_i^{sc} < 
\nu_x < \nu_c^{sc}$, for which $\beta_x^{fs} = p/2$, $\beta_x^{sc} =(p-1)/2$
and
\beq
 \alpha_{fs} = \frac{1}{4} (6\beta_x + 1) 
\eeq
independent of the circumburst medium stratification, where $\beta_x \equiv (p-1)/2$
(i.e. $\beta_x$ is the slope of the up-scattered spectrum, which is what XRT would 
measure if the scattered emission were brighter than forward-shock's). 
The ratio of intensities in equation (\ref{Fnu}) is $(\Gamma/\gamma)^{2\beta_x} 
(\nu/\gamma\nu'_c)^{1/2}$, leading to the same light-curves of the scattered emission 
as given in equations (\ref{F1}) and (\ref{F0}), with $\beta_0$ of equation (\ref{b01}).

\subsection{ $\nu'_i, \nu'_c < \nu_x\gamma/\Gamma^2 < \nu_x/\gamma$ }
\label{case3}

 This is the $\nu_i^{fs},\nu_c^{fs} < \nu_x$ and $\nu_i^{sc},\nu_c^{sc} < \nu_x$
case, for which $\beta_x^{fs} = \beta_x^{sc} = p/2$ and
\beq
 \alpha_{fs} = \frac{1}{2} (3\beta_x - 1)
\label{ncnx}
\eeq
with $\beta_x \equiv p/2$. 
The intensity ratio of equation (\ref{Fnu}) is $(\Gamma/\gamma)^{2\beta_x}$,
leading to a scattered flux as given in equation (\ref{int}) with the exponent 
\beq
 A = 4 - 1.5s - \beta_x (3-0.5s) 
\eeq
thus 
\beq
  \beta_0 = \frac{8-3s}{6-s} = \left\{ \begin{array}{ll}
          4/3 & s=0 \\ 1/2 & s=2 \end{array}  \right.
\label{b02}
\eeq
For $\beta_x < \beta_0$, the light-curve of the scattered emission is
\beq
 F_\nu^{sc}(T) \propto \frac{dM}{dT}  \Gamma^{2\beta_x} \left\{ \begin{array}{ll} 
           T^{-(3/4)\beta_x+1} & s=0 \\ T^{-\beta_x+1/2} & s=2 \end{array} \right. \;.
\label{F2}
\eeq
For $\beta_x > \beta_0$, the scattered emission satisfies equation (\ref{F0}).
                                                                                                    
\subsection{ $\nu'_c < \nu_x\gamma/\Gamma^2 < \nu'_i < \nu_x/\gamma$ }

 In this case ($\nu_i^{fs},\nu_c^{fs} < \nu_x$ and $\nu_c^{sc} < \nu_x < \nu_i^{sc}$),
we have $\beta_x^{fs} = p/2$, $\beta_x^{sc} = 1/2$, and $\alpha_{fs}$ is given by 
equation (\ref{ncnx}). The intensity ratio in equation (\ref{Fnu}) is $(\Gamma/\gamma)
(\nu/\gamma\nu'_i)^{\beta_x-1/2}$, leading to
\beq
 F_\nu^{sc}(T) \propto \frac{dM}{dT} \Gamma \left\{ \begin{array}{ll}
           T^{5/8} & s=0 \\ T^0 & s=2 \end{array} \right. \;.
\label{F3}
\eeq

\section{Scattered emission -- numerical results}
\label{numerical}

\begin{figure*}
\centerline{\psfig{figure=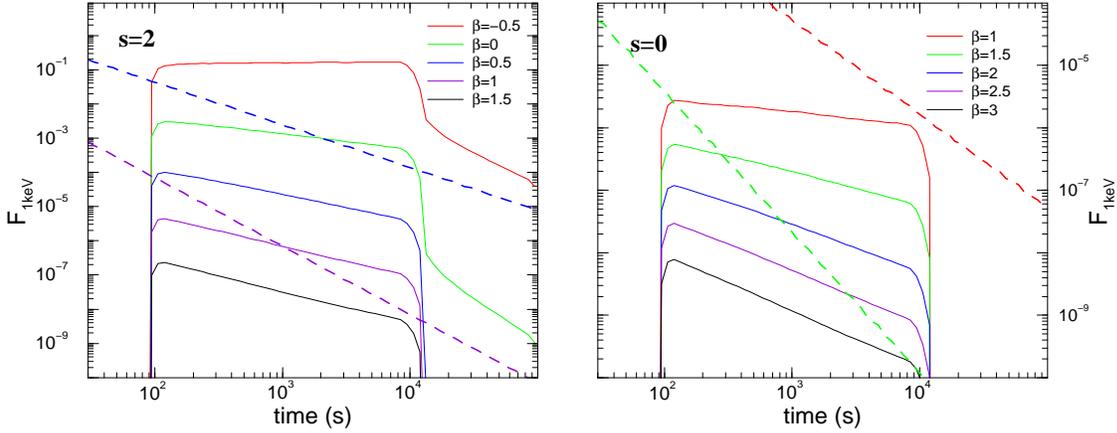,width=15cm}}
\caption{ X-ray light-curves (flux in arbitrary units) from the forward-shock (thick
  lines) and a radially-extended scattering outflow (thin lines), for various slopes 
  $\beta$ of the forward-shock emission spectrum.
  The dynamics of this shock is given by $E_{fs}/A_*=10^{53}$ erg in the left panel
  (for a wind-like medium, $s=2$) and  $E_{fs}/n=10^{53} {\rm erg\,cm^3}$ in the
  right panel (for a homogeneous circumburst medium, $s=0$). 
  The cooling frequency of the forward-shock synchrotron emission is set above X-rays 
  and injection frequency is $h\nu_i (1s)= 1$ eV. The brightness of the forward-shock
  emission decreases strongly with increasing spectral slope.
  The scattering outflow has $E_{sc}=10^{53}$ erg, a constant Lorentz factor $\Gamma$ 
  ($10^3$ for left panel, $10^4$ for right panel), constant mass ejection rate $dM/dt_{lag}$,
  is purely leptonic, and was ejected from 100 s to 10 ks after the forward-shock
  (hence the scattered emission arrives at observer at 0.1--10 ks).
  Some of the spectral slopes are unrealistically soft or hard -- XRT measures
  $\beta_x \in (0.5,1.5)$ -- but have been used to show that the decay of the scattered 
  emission depends on $\beta$ if the forward-shock emission is hard,
  becoming independent of $\beta$ if the forward-shock emission is soft.
  The transition between these two regimes takes place at $\beta_0 \simeq 0.5$
  for $s=2$ and at $\beta_0 \simeq 2$ for $s=0$. The scattered emission shown in 
  the left panel after 10 ks (for the two hardest spectra) is the "large-angle 
  emission", arriving from the fluid moving at angle $\gg \Gamma^{-1}$ relative
  to the direction toward the observer.
   }
\label{betalc}
\end{figure*}

 The above analytical derivations for the scattered emission light-curve were made
for photons moving along the explosion center -- observer axis, for which equation
(\ref{tfs}) can be solved, yielding the trivial solution of equation (\ref{tfstsc}),
$t_{fs} \simeq t_{sc}$. For other photons ($\alpha' > 0$), the Doppler factor of the
scatterer--forward-shock relative motion is of the same order, $D \simeq 
\Gamma/\gamma(t_{fs})$, but the retarded emission time cannot be obtained analytically.

 The scattered emission from a radially-extended outflow is calculated numerically by 
discretizing it into surfaces and by using the relevant equations of \S\ref{setup}--
\S\ref{analytical} to calculate the emission from each surface. We have verified that 
(1) the decays given in equations (\ref{F1}) and (\ref{F2}) (obtained in the $\alpha'=0$ 
approximation) are correct and 
(2) the decay of the scattered light-curve is independent of the incident spectrum 
for softer spectra (i.e. independent of the decay of the forward-shock emission). 
Therefore, as expected analytically, there is a "critical" spectral slope $\beta_0$ of 
the forward-shock emission such that the scattered flux received at some time $T$ is 
mostly that produced by the forward shock at same time $T$, if $\beta < \beta_0$, or
at the onset of deceleration, if $\beta > \beta_0$. 
However, the numerical integration of the scattered flux leads to a $\beta_0$ that is 
larger than that given in equations (\ref{b01}) and (\ref{b02}) by $\sim 1$. 

 For the case presented in \S\ref{case1}, we find numerically that $\beta_0 \simeq 2$ 
for $s=0$ and $\beta_0 \simeq 0.5$ for $s=2$ (the dependence of the scattered emission 
decay on the incident spectrum is shown in Figure \ref{betalc}), while for that of 
\S\ref{case3}, $\beta_0 \simeq 2.5$ for $s=0$ and $\beta_0 \simeq 1.5$ for $s=2$. 
As the X-ray spectral slopes measured by XRT are between 0.5 and 1.5, the decay
of the X-ray plateau resulting from scattering the forward-shock emission should be 
independent of $\beta_x$ only if (1) the circumburst medium is a wind, (2) $\nu_i^{fs} 
< \nu_x < \nu_c^{fs}$, and (3) $\nu_i^{sc} < \nu_x < \nu_c^{sc}$. In all other cases, 
the decay of the scattered emission is correlated with the spectral slope $\beta_x$.

 For the parameters used in Figure \ref{betalc}, the scattered emission is brighter 
than the forward-shock's (and yields a light-curve plateau) only if the spectrum of 
the forward-shock emission is softer than $\beta \simeq 1.2$ for $s=0$ and $\beta 
\simeq 0.8$ for $s=2$. Numerically, we find that X-ray plateaus can be obtained for 
harder forward-shock emissions (down to the hardest observed XRT spectra -- $\beta_x 
\simeq 0.5$) if (1) the cooling frequency of the forward-shock emission is below X-rays
and the up-scattered cooling frequency above, (2) the Lorentz factor or energy of the 
scattering outflow is larger, or (3) the scattering electrons are relativistic (i.e. 
they inverse-Compton scatter the incoming photons).

\subsection{Chromatic plateaus and breaks}

\begin{figure*}
\centerline{\psfig{figure=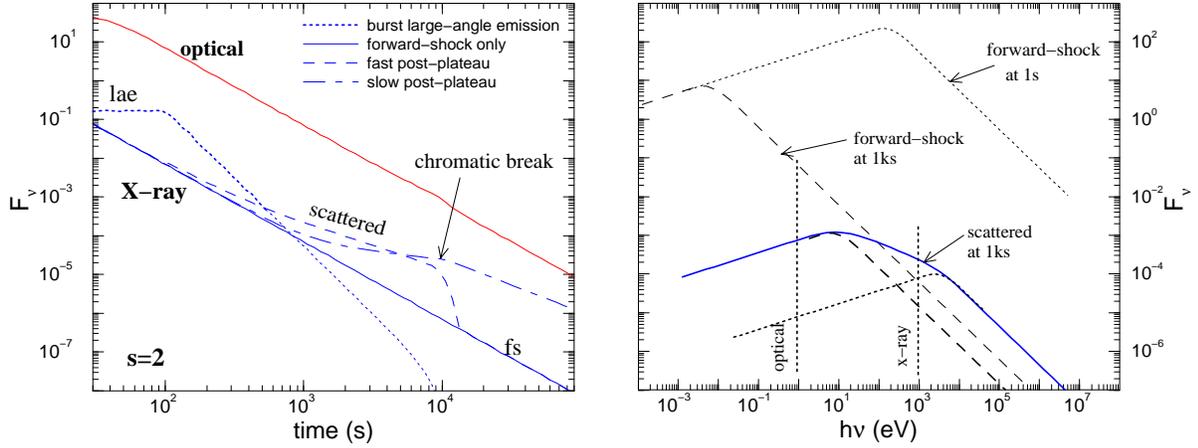,width=16cm}}
\caption{Light-curves and spectra of forward-shock and scattered emissions.
     {\it Right panel}: spectrum of synchrotron forward-shock emission at 1 s and 1 ks 
      and spectrum of up-scattered emission at 1 ks.
      Thick lines show the 1s (dotted) and 1 ks (dashed) spectra of the up-scattered 
      forward-shock emissions, shifted vertically toward the integrated spectrum to 
      match their expected peak frequencies.
      Note that the scattered emission is brighter than the forward-shock's in the X-ray
      but not in the optical, which leads to a decoupling of the two light-curves and
      to chromatic X-ray breaks. Also, note that the scattered emission may be harder 
      in the X-rays than the underlying forward-shock spectrum.
     {\it Left panel}: chromatic X-ray light-curve breaks produced by the scattered 
      emission. At early times, the X-ray afterglow emission is dominated by the
      "large-angle" prompt emission released during the burst but arriving later at
      observer (owing to the spherical curvature of the emitting surface). 
      The decay of the X-ray emission reflects the distribution of mass $dM/dt_{lag}$ and 
      Lorentz factor $\Gamma$ in the scattering outflow with geometrical depth $ct_{lag}$
      (distance from forward-shock): if either $dM/dt_{lag}$ or $\Gamma(t_{lag})$ fall 
      sharply at some distance $ct_{break}$ then the X-ray break at observer time 
      $t_{break}$ will be very sharp (dashed line), as observed for a minority of 
      Swift afterglows; a more gradual decrease of $dM/dt_{lag}$ or $\Gamma(t_{lag})$ 
      at $t_{lag} > t_{break}$ yields a steepening of the X-ray decay at $t_{break}$ 
      (dot-dashed curve), as displayed by a majority of Swift afterglows.
      {\sl Parameters}. Forward-shock: $E_{fs}/A_*=10^{53}$ erg (wind-like medium); 
      $\beta = 1$, $h\nu_i (1s)= 100$ eV, $\nu_c$ above X-rays. 
      Scattering outflow: uniform $\Gamma=10^3$, baryon-free composition, 
      $t_{break} = 10$ ks, and 
      (1) $E_{sc}=10^{53}$ erg, $dM/dt_{lag} = 0$, cold outflow, ending at $t_{break}$
      for the X-ray light-curve with a sharp plateau end, 
      (2) $E_{sc}=10^{54}$ erg, $dM/dt_{lag} \propto t_{lag}^{1/2}$ for $t_{lag}<t_{break}$, 
      $dM/dt_{lag} \propto t_{lag}^{-1/2}$ for $t_{lag}>t_{break}$, hot outflow with 
      $\gamma_e = 10$ for the X-ray light-curve with a slow post-plateau decay. 
      The optical flux is the same for both scattering outflow parameters because 
      the scattered emission is dimmer than the forward-shock's.
 }
\label{plateau}
\end{figure*}

 As illustrated in Figure \ref{plateau}, the up-scattered emission may overshine 
the forward-shock's only at higher photon energies (in the X-rays) but not at lower
frequencies (e.g. in the optical). This implies that the X-ray plateau produced by
scattered forward-shock emission does not appear in the optical. Consequently, the 
end of the plateau will also be a chromatic feature. Such chromatic light-curve breaks 
are observed for several GRB afterglows: GRB 050401 (Watson et al 2006 - Figure \ref{xlc}); 
GRBs 050319, 050607, 050713A, 050802, 050922C (Panaitescu et al 2006b); GRB 070110 
(Troja et al 2007 - Figure \ref{xlc}), GRB 050318 (Liang et al 2007) and, as noted
before, cannot be accommodated by the forward-shock emission alone. 

\subsection{Diversity of plateau decays and flare shapes}
\label{shapes}

\begin{figure}
\psfig{figure=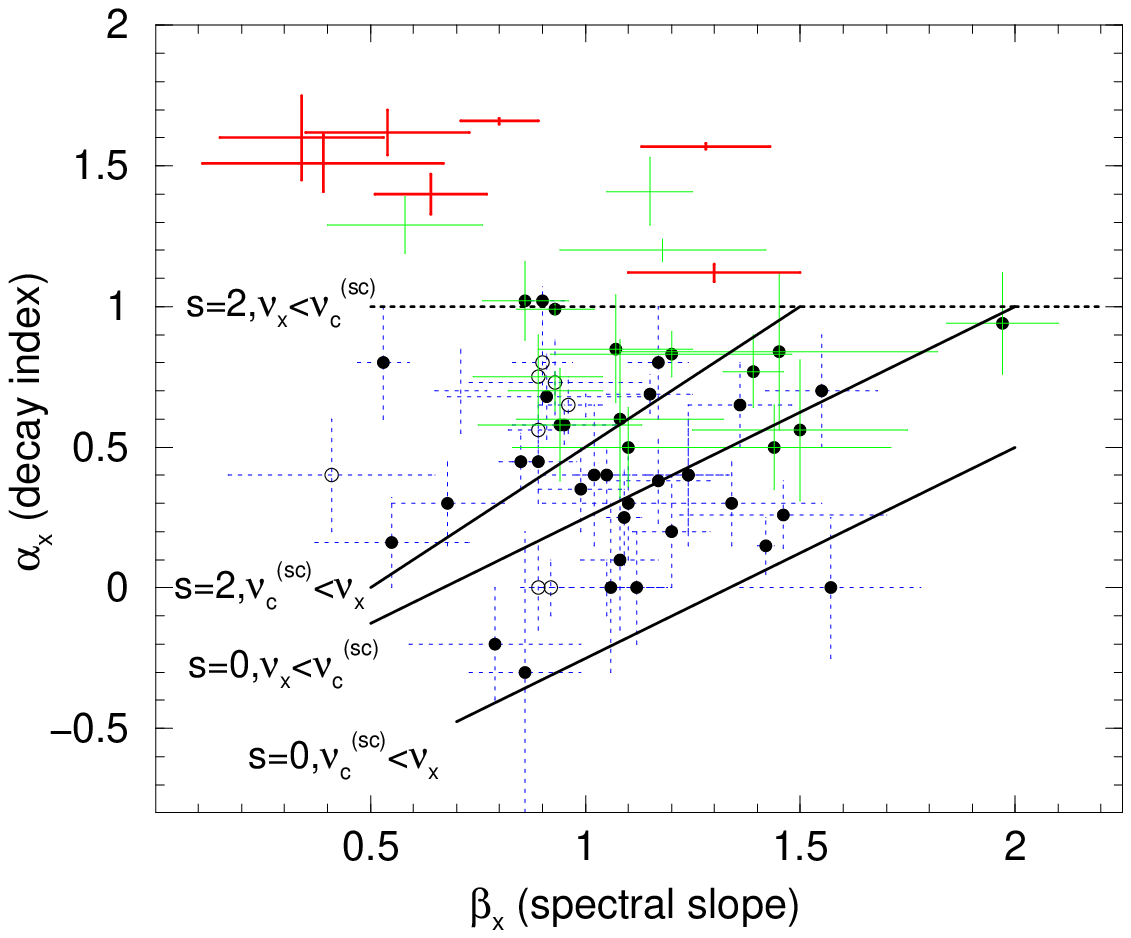,width=8cm}
\caption{Power-law decay index of the X-ray plateau ($F_\nu \propto T^{-\alpha_x}$)
     vs. spectral slope ($F_\nu \propto \nu^{-\beta_x}$) for 40 Swift afterglows with 
     flattening followed by a steepening of the X-ray light-curve at 1--30 ks 
     (blue, dotted error bars). Also shown are 7 afterglows with a single power-law 
     X-ray light-curve decay after burst and until 10--100 ks (red, thick error bars), 
     and 17 afterglows with a flattening of the X-ray light-curve after burst tail but 
     without a steepening observed until last observation, days after trigger (green, 
     thin error bars).
     Lines show the model expectations for the scattered forward-shock emission model
     (equations \ref{F1} and \ref{F2}) for a constant mass-flux ($d^2M/dt_{lag}^2=0$) and
     Lorentz factor ($d\Gamma/dt_{lag}=0$) of the scattering outflow. The range of model 
     decay indices expected is comparable to that observed for the 57 afterglows with
     an X-ray plateau (whether or not the plateau end was observed).
     75 percent (filled symbols) and 90 percent (all symbols) of afterglows are consistent 
     within $1\sigma$ and $2\sigma$, respectively, with this model, thus 10--25 percent 
     of afterglows require scattering outflows with non-constant $dM/dt_{lag}$ and/or 
     $\Gamma(t_{lag})$. 
     Note that the 7 X-ray afterglows without slow decays are, on average, harder than 
     the 57 afterglows with slow fall-offs, as expected for the scattering-outflow model.
    }
\label{a2b2}
\end{figure}

 In Figure \ref{a2b2}, we compare the light-curve decay indices and spectral slopes 
measured by XRT during the X-ray plateau with the analytical results of equations 
(\ref{F1}) and (\ref{F2}) for a constant mass-flux $dM/dt_{lag}$ and Lorentz factor 
$\Gamma$ of the scattering outflow, using the $\beta_0$ derived numerically (instead 
of that given in equations \ref{b01} and \ref{b02} for the central line-of-sight photons). 

 The spectral slopes displayed in Figure \ref{a2b2} show that the afterglows without 
plateaus (i.e. those for which the forward-shock emission is brighter than the scattered 
emission at all times) are generally harder than the afterglows with plateaus.
This confirms the expectation illustrated in Figure \ref{betalc} for the scattering-outflow
model, that plateaus are easier to obtain if the forward-shock emission is softer 
(as this leads to a faster-decaying forward-shock light-curve).

 Although most of the afterglow decays displayed in Figure \ref{a2b2} are in the 
$\alpha_x < 1$ region, where the decay index $\alpha_x$ depends on the spectral 
slope $\beta_x$, these two quantities are not correlated (linear correlation 
coefficient $r(\alpha_x,\beta_x)= 0.03 \pm 0.10$). This can be due either to that all  
cases analyzed in \S\ref{case1}--\S\ref{case3} occur in reality or to that $dM/dt_{lag}$ 
and $\Gamma$ vary with ejection time $t_{lag}$ (which is measured directly by the observer
time) for some/most afterglows. 

 For a power-law radial distribution of the scattering outflow mass-flux ($dM/dt_{lag} 
\propto t_{lag}^{-m}$) and Lorentz factor ($\Gamma \propto t_{lag}^{-g}$, with $g \geq 0$ 
because internal shocks lead to a Lorentz factor decreasing outward), the decay index 
$\alpha_{sc}$ of the scattered emission is 
\beq
 \alpha_{sc} = m + 2g\beta_x + \alpha_{sc}(m=0,g=0)
\label{mg}
\eeq
with $\alpha_{sc}(m=0,g=0)$ being the decay index given in equations (\ref{F1}), 
(\ref{F0}), (\ref{F2}), and (\ref{F3}) for a uniform outflow.

\begin{figure}
\psfig{figure=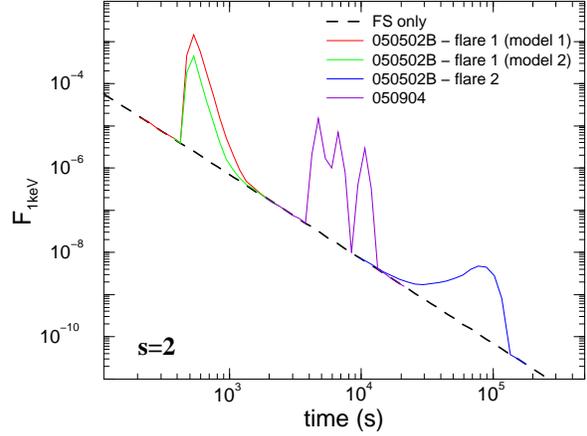,width=8cm}
\caption{Flaring episodes for the scattered forward-shock emission obtained with fast 
   evolving functions for the mass-flux $dM/dt_{lag}$ and Lorentz $\Gamma(t_{lag})$,
   chosen so that the X-ray flares have a morphology resembling those of the flares
   of the afterglows indicated in legend (see Figure \ref{xlc}). The lab-frame timing 
   of the scattering outflow ejection can be read directly from the observer time. 
   General parameters: $E_{fs}/A_*=10^{53}$ erg (wind-like medium), $\beta=1$, 
   $h\nu_i (1s)= 1$ eV, $\nu_c$ above X-rays; $E_{sc}=10^{53}$ erg, baryon-free composition. 
   Other parameters: 
   (1) model 1 for first flare of GRB 050502B -- 
       $d^2M/dt_{lag}^2 = 0$, $\Gamma= 10^4 (t_{lag}/500s)^{-5}$;  
   (2) model 2 for first flare of GRB 050502B -- 
       $dM/dt_{lag} \propto t_{lag}^{-10}$, $\Gamma= 10^4$ (constant);
   (3) second flare of GRB 050502B -- $dM/dt_{lag} \propto t_{lag}^3$, $\Gamma= 3000$;
   (4) GRB 050904 -- $dM/dt_{lag} \propto t_{lag}^{-1}$, $\Gamma= 10^4$ 
  (3 episodes of ejection).
 }
\label{eg}
\end{figure}

 Therefore, a variety of X-ray plateau decays can be obtained by varying the indices
$m$ and $g$ of equation (\ref{mg}). The same applies to the shape of the flares resulting 
from up-scattering the forward-shock emission (Figure \ref{eg}).
In particular, decreasing $dM/dt_{lag}$ or $\Gamma(t_{lag})$ yield flares with a slow 
fall, while an increasing $dM/dt_{lag}$ produces flares with a slow rise, which can 
account for the diverse morphology of X-ray flares identified by Chincarini et al (2007).

\subsection{Fast post-plateau decays}
\label{fastdecay}

 Given that the post-plateau X-ray decay depends on the $dM/dt_{lag}$ and $\Gamma(t_{lag})$ 
distributions at $t_{lag} > t_{break}$, the scattering outflow model can account 
(Figure \ref{plateau}) for the extremely fast-decaying $F_x \propto T^{-8\pm2}$ 
post-plateau emission observed for GRB afterglow 070110 (Troja et al 2007) after 20 ks 
and the $F_\nu \propto T^{-3}$ post-plateau decays of GRBs 050730 (Perri et al 2007), 
060413, and 060607A (Molinari et al 2007, Nysewander et al 2007) after 10 ks, and of 
GRB 070311 (Guidorzi et al 2007) after 150 ks. 
If the scattering outflow is concentrated in thin shells then the same model can account
for the fast decay of the afterglow emission after flaring episodes, followed by a smooth
and slower decay (from the emergent forward-shock emission), as observed for GRB afterglow 
051117A (Goad et al 2007) at 10 ks.
Such fast decays cannot be explained by the standard forward-shock model.

 After the X-ray plateau, the majority of afterglows exhibit a light-curve decay $F_x 
\propto t^{-\alpha}$ with $\alpha \in (1,2)$. This indicates that $dM/dt_{lag}$ and
$\Gamma(t_{lag})$ are nearly uniform distributed (i.e. the indices $m$ and $g$ of 
equation \ref{mg} are not far from zero) and that large values of these parameters, 
mimicking a sharp cut-off in their distribution and yielding sharp post-plateau decays, 
are rare. Figure \ref{plateau} (left panel) illustrates the chromatic X-ray break, 
followed by a slow post-plateau decay, that is obtained with a broken power-law for 
$dM/dt_{lag}$, the transition between the two power-laws being made at the epoch of 
the X-ray break.

\subsection{Spectral evolution of flares and plateaus}

 Figure \ref{plateau} (right panel) also shows that scattering of the forward-shock 
emission may yield a spectrum with a turnover extending up to (or above) the X-ray 
range, thus the scattered emission may be harder than the intrinsic forward-shock 
emission. This implies that, sometimes, X-ray plateaus and flares could be harder 
than the preceding and following X-ray emission. A spectral softening at the end
of plateau is not observed (in general), which indicates that either the spectral 
turn-over formed by up-scattering the peak frequency of the synchrotron forward-shock 
emission lies below the X-ray or that it is in the X-rays but does not evolve. 
In the latter case, the break at the end of plateau should be attributed to a change 
in the distribution of outflowing mass with depth, as the alternative explanation
-- a decreasing outflow Lorentz factor  -- would produce a softening of the peak 
energy of the up-scattered spectrum (equation \ref{boost}).

 A hardening of the X-ray spectrum, followed by softening, is observed for the flares
of GRB 050502B (Falcone et al 2006) and, perhaps, those of GRB 050904 (Cusumano et al 
2007) occurring {\it during the afterglow phase}, which suggests that the turn-over 
at the peak of the up-scattered emission spectrum reaches sometimes the 0.3--10 keV 
range. This is consistent with flares originating from regions in the scattering
outflow of Lorentz factor larger than the average (i.e. than for the part of the
outflow yielding the plateau), although it is possible that the spectral peak energy 
of the flaring emission is larger than that of the plateau's simply because flares 
occur at earlier times, when the peak frequency of the seed forward-shock emission 
is higher.

\subsection {Cold and hot scattering outflows} 

 If the scattering electrons are relativistic (with random Lorentz factor 
$\gamma_e$ in the scattering outflow's frame), they will inverse-Compton scatter the
incident photon and increase its energy by a factor $\simeq \gamma_e^2$ (in the frame
of the scattering fluid). Then, the comoving intensity in the numerator of the $rhs$
of equation (\ref{frac1}) is calculated at comoving frequency $\nu\gamma/(\Gamma\gamma_e)^2$. 
If this frequency is above the peak frequency of the synchrotron forward-shock emission 
(in the shock's frame), then the scattered flux will be a factor $\gamma_e^{2\beta_x}$ 
larger than that obtained for a cold outflow (equations \ref{F1}, \ref{F0}, and \ref{F2}).

 Hence, the decrease of the scattered flux by factor $m_p/m_e = 1837$ resulting from
going from a cold and purely-leptonic scattering outflow (as considered so far) to one 
with a unity baryon-to-lepton ratio can be compensated for if the latter is hot, so that 
the incident photons are inverse-Compton scattered off electrons with $\gamma_e = 
(m_p/m_e)^ {0.5/\beta_x}$. This holds true as long as all incident photons up-scattered 
to observer frequency $\nu$ are (before being scattered) above the peak frequency of the 
forward-shock synchrotron spectrum.
 For $\beta_x > 1/2$ (as observed), it follows that $\gamma_e < m_p/m_e$, which implies 
that the internal shocks energizing the scattering outflow need not be relativistic and 
the necessary $\gamma_e$ could be produced by small fluctuations in the Lorentz factor of 
the scattering outflow.

 Alternatively, that $F_{sc}(\nu) \propto (\Gamma \gamma_e)^{2\beta_x}$ 
implies that a hot, scattering outflow of bulk Lorentz factor $\Gamma/\gamma_e$ and 
comoving-frame electron energy $\gamma_e m_e c^2$ yields the same X-ray scattered flux 
as a cold outflow of Lorentz factor $\Gamma$. Thus, if the scattering outflow is hot, 
then an X-ray plateau can be obtained for a Lorentz factor $\Gamma$ smaller than that
considered for Figures \ref{flares}, \ref{betalc}, \ref{plateau}, and \ref{eg}. 

\subsection{Circumburst medium}
                                                                                                
 For a given scatterer Lorentz factor $\Gamma$ and forward-shock spectral slope $\beta$, 
a wind-like circumburst medium is more likely to yield an X-ray scattered emission 
brighter than the forward-shock's. This is primarily due to that, at a fixed observer
time, the Lorentz factor of the forward-shock is lower for a wind-like medium than
for a homogeneous one, as shown by equations (\ref{g0}) and (\ref{g2}), hence the
scattering has a stronger effect (quantified by the ratio $\Gamma/\gamma$) for the 
former case. Comparable forward-shock Lorentz factors would require a much higher
homogeneous medium density ($n \simeq 10^4\, {\rm cm^{-3}}$), which is at odds with
with the low synchrotron self-absorption frequency (below 10 GHz), or a much
lower wind density ($A_* \simeq 0.01$), which is $>30$ times less tenuous than measured 
for Galactic WR stars. 

 A second reason for which the scattered emission is more prominent
for a wind-like medium is at work when the cooling frequency of the forward-shock
emission is above X-rays: in this case, the forward-shock flux decreases faster for
a wind-like medium, allowing the scattered emission to emerge. Therefore, the 
scattering-outflow model for the X-ray plateaus favours a circumburst medium with the 
$r^{-2}$ radial stratification expected for a massive star as the GRB progenitor. 

 However, we note that, if the scattering electrons are relativistic, then the ratio 
of scattered and direct fluxes is quantified by $\Gamma \gamma_e/\gamma$, thus a 
hotter scattering outflow could compensate for the above two factors which favour a 
wind-like medium.

\section{Conclusions}

 The results presented in preceding sections illustrate that the up-scattering of the 
forward-shock emission by a more relativistic, lepton-enhanced outflow can accommodate 
four puzzling X-ray features discovered by Swift: flares, plateaus, chromatic light-curve 
breaks, and sharp post-plateau decays. 
 The model requires the existence of a long-lived engine that releases relativistic
ejecta for at least 10 ks or some other way of producing a relativistic outflow that 
has such a radial spread. That the scattering outflow has a higher Lorentz factor 
than the forward shock need not be a feature of the engine and could be just the 
consequence of the deceleration that the forward shock undergoes as it sweeps-up and 
energizes the ambient medium. The requirement that the scattering outflow is pair-rich
is to ensure a sufficiently large (sub-unity) optical thickness to electron scattering, 
so that the scattered emission can overshine the forward-shock's direct emission and 
explain the above three features of X-ray afterglows. This requirement can be relaxed 
if the scattering outflow is hot (by birth or due to internal shocks), so that the 
incident photons are also inverse-Compton scattered (in addition to their bulk up-scattering). 

 Of the above-mentioned X-ray features, only plateaus can be explained by the
forward-shock model: energy injection in this shock by means of some delayed ejecta 
should lead to a slower decay of the forward-shock emission. However, the other three
features (flares, chromatic breaks, and sharp post-plateau fall-offs) cannot be accounted
for by the forward-shock model, where fluctuation timescales are expected to be of
order of the afterglow age and the multiwavelength light-curve behaviours should be 
well coupled.

\subsection{X-ray flares} 

 The light-curves for the scattered forward-shock emission shown in Figure (\ref{eg}) 
show that this model can account for the sharpness and brightness of the flares observed 
in many X-ray afterglows. The dynamics of sweeping-up
the photons left behind by the forward shock leads to flares of duration $\delta T =
(\gamma/\Gamma)^2 T$, with $\gamma$ and $\Gamma$ the forward-shock Lorentz factor
at observer time $T$ and $\Gamma$ that of the scatterer. This $\delta T$ is also the
spread in the photon-arrival time caused by the spherical curvature of the scattering
surface: $\delta T \simeq R/\Gamma^2 \simeq (\gamma^2 T)/\Gamma^2)$, where $R \simeq
\gamma^2 T$ is the radius of the scatterer when it catches-up with the forward shock.
Therefore, very short flares are a natural feature of the scattered emission, provided
that the radial spread of the scattering outflow is sufficiently small (instantaneous
ejection). For longer-lived flares, the shape of the flare (fast-rise, fast-fall, or 
symmetric flares) depends on the radial distribution of the mass-flux $dM/dt_{lag}$ and/or 
of the Lorentz factor $\Gamma(t_{lag})$ in the scattering outflow (Figure \ref{eg}).

\subsection{X-ray plateaus, chromatic breaks, and sharp post-plateau decays}

 Figure \ref{betalc} illustrates that X-ray plateaus (as observed for most afterglows) 
can result from scattering the forward-shock emission. Because the up-scattered 
emission is more likely to overshine the direct forward-shock emission at higher photon 
energies than at lower energies, this model can also produce chromatic plateaus and 
breaks, which appear in the X-rays but not in the optical as well (Figure \ref{plateau}). 
Such chromatic X-ray breaks are observed for several GRB afterglows.

 Sharp decays following X-ray plateaus, as observed for a few GRB afterglows, result if 
the $dM/dt_{lag}$ or $\Gamma(t_{lag})$ of the scattering outflow decrease sufficiently 
fast at some point behind the forward shock (i.e. the scattering outflow has a well-defined 
trailing edge). Only a few XRT afterglows display a sharp post-plateau decay, the norm 
being that of a smooth transition to a steeper power-law decay. This requires scattering 
outflows with a more complex structure, where the plateau end corresponds to a change 
in the radial distribution of $dM/dt_{lag}$ and/or $\Gamma(t_{lag})$. That most post-plateau
X-ray light-curves have a slow decay indicates that the radial distributions of $dM/dt_{lag}$ 
and $\Gamma(t_{lag})$ (equation \ref{mg}) are not far from being uniform and that they
rarely exhibit the fast decrease with distance from the forward-shock that is required 
by the minority of X-ray afterglows with sharp post-plateau decays.

\subsection{Prolonged activity of central engine}

 For a scattering outflow of Lorentz factor well above that of the forward shock, the 
arrival time of the scattered photons is nearly equal to the delay (in the lab frame) 
between the ejection of the forward-shock driving ejecta and the scattering fluid. 
This has two consequences. {\it First}, in addition to the spectral slope of the 
forward-shock 
emission, the scattered flux received at an observer time $T$ reflects the properties 
($dM/dt_{lag}$ and $\Gamma(t_{lag})$) of the scattering fluid ejected at lab-frame time 
$t_{lag}=T$. This implies that a variety of X-ray plateau decays and shapes can be 
obtained by choosing the right functions for $dM/dt_{lag}$ and $\Gamma(t_{lag})$
(equations \ref{F1}, \ref{F0}, \ref{F2}, and \ref{F3}). The {\it second} is that a flare 
seen at time $T$ or a plateau lasting until $T$ require that the central engine operates 
at/until a lab-frame time equal to $T$. Thus, the scattering-outflow model for X-ray 
flares and plateaus is still based on the existence of a long-lived engine, as is the 
internal-shock interpretation of flares.

\subsection{Scattering outflow vs forward shock}

 The post-plateau decay of Swift X-ray afterglows is generally compatible with a forward-shock
origin (see fig. 5 of Willingale et al 2007, fig. 11 of Panaitescu 2007, fig.  6 of Liang et 
al 2007). However, no single variant of that model that can account for the spread of
post-plateau decay indices for a given spectral slope. Furthermore, for a set of 60 Swift
afterglows, we find that the post-plateau X-ray decay index is not correlated with the 
spectral slope (linear correlation coefficient $r = -0.18 \pm 0.10$), contrary to what is 
expected for the forward-shock emission. 

 That lack of correlation is a natural consequence of the scattering model, where the 
post-plateau decay is determined not only by the spectral slope of the forward-shock seed 
photons but also by the radial distribution of $dM/dt_{lag}$ and/or $\Gamma(t_{lag})$
in the scattering outflow. A stronger argument in favour of this model for X-ray afterglows
is that it can explain two features (chromatic light-curve breaks and sharp post-plateau 
decays) that cannot be accounted for by the forward-shock model. 

 However, about 30 percent of the well-monitored X-ray afterglows of Willingale (2007)
do not have a plateau, thus their emission need not be attributed to a scattering outflow, 
and could originate in the forward shock. Furthermore, the GRB afterglows 051109A (Yost et al 
2007), 060614 (Mangano et al 2007a), 060714 (Krimm et al 2007), and 060729 (Grupe et al 2007) 
display an achromatic break, which is seen simultaneously in the optical and X-ray
light-curves. Such breaks require that both emissions arise from the same mechanism.
Although that could be achieved by the scattering outflow model, provided that the 
forward-shock emission peaks at a sufficiently low energy that the scattered flux is
dominant also in the optical, it seems more natural to attribute the achromatic breaks at 
plateau ends to the cessation of energy injection into the forward shock. 

 The forward-shock is also a more plausible origin for the long-lived,  power-law 
X-ray afterglows of e.g. GRB 050416A (Mangano et al 2007b), GRB 050822 (Godet et al 2007), 
GRB 060319 (Burrows \& Racusin 2007) GRB 060729, which, in the scattering outflow model, 
would require a central engine producing a scattering outflow with a nearly uniform 
distribution of $dM/dt_{lag}$ and $\Gamma$ over very long times ($\siml 1$ Ms in source frame). 

 Therefore, the features of optical and X-ray afterglow light-curves provide evidence in 
favour of both the forward-shock and scattering-outflow models. When the forward-shock
emission is the brightest, the X-ray afterglow may lack a plateau or, if it has one owing 
to energy injection in the blast-wave, then the plateau ends with an achromatic break.
If the scattered emission is dominant, then the X-ray afterglow exhibits a plateau, most 
likely ending with a chromatic break. This picture is supported by that X-ray afterglows 
without plateaus are harder (Figure \ref{a2b2}), as a harder forward-shock emission is 
expected to be brighter and, thus, more likely to overshine the scattered emission.

 Thus, the observed diversity of Swift X-ray afterglows is due to both mechanisms being at 
work, in addition to the diversity that each model can produce on its own.

\section*{Acknowledgments}
 This work was supported by NASA Swift Guest Investigator grant NNG06EN00I



\begin{thebibliography}{99}
\bibitem{} Burrows D. et al, 2007, Phil Trans A, 365, 1213 
\bibitem{} Burrows D., Racusin J., 2007, Nuovo Cimento B, 121, 1273  
\bibitem{} Chincarini G. et al, 2007, ApJ, submitted (astro-ph/0702371)
\bibitem{} Cusumano G. et al, 2007, A\&A, 462, 73 
\bibitem{} Falcone A. et al, 2006, ApJ, 641, 1010
\bibitem{} Fan Y., Piran T., 2006, MNRAS, 197, 206
\bibitem{} Fenimore E., Ramirez-Ruiz E., 1999, preprint (astro-ph/9909299)
\bibitem{} Genet F., Daigne F., Mochkovitch R., 2007, MNRAS, 381, 732
\bibitem{} Goad M. et al, 2007, A\&A, 468, 103            
\bibitem{} Godet O. et al, 2007, A\&A, 471, 385           
\bibitem{} Grupe D. et al, 2007, ApJ, 662, 443
\bibitem{} Guidorzi C. et al, 2007, A\&A, accepted (arXiv:0708.1383)
\bibitem{} Ioka K. et al, 2006, A\&A, 458, 7
\bibitem{} Krimm H. et al, 2007, ApJ, 665, 554
\bibitem{} Kumar P., Panaitescu A., 2000, ApJ, 541, L51
\bibitem{} Kumar P. et al, 2007, MNRAS, 367, L52
\bibitem{} Liang E., Zhang B-B., Zhang B., 2007, ApJ, accepted \\ (arXiv:0705.1373)
\bibitem{} Lyutikov M., 2006, New J. of Phys., 8, 119 
\bibitem{} Mangano V. et al, 2007a, A\&A, 470, 105
\bibitem{} Mangano V. et al, 2007b, ApJ, 654, 403
\bibitem{} M\'esz\'aros P., Rees M., 1997, ApJ, 476, 232
\bibitem{} Molinari E. et al, 2007, A\&A, 469, L14
\bibitem{} Nousek J. et al, 2006, ApJ, 642, 389
\bibitem{} Nysewander M. et al, 2007, ApJ, submitted (arXiv:0708.3444)
\bibitem{} Paczy\'nski B., Rhoads J., 1993, ApJ, 418, L5
\bibitem{} Panaitescu A. et al, 2006a, MNRAS, 366, 1357
\bibitem{} Panaitescu A. et al, 2006b, MNRAS, 369, 2059
\bibitem{} Panaitescu A. et al, 2007, MNRAS, 379, 331
\bibitem{} Perri M. et al, 2007, A\&A, 471, 83
\bibitem{} Ramirez-Ruiz E., Merloni A., Rees M., 2001, MNRAS, 324, 1147 
\bibitem{} Rees M., M\'esz\'aros P., 1994,  ApJ, 430, L93
\bibitem{} Sari R., Piran T., Narayan R., 1998, ApJ, 497, L17
\bibitem{} Troja E. et al, 2007, ApJ, 665, 599
\bibitem{} Uhm Z., Beloborodov A., 2007, ApJ, 665, L93
\bibitem{} Watson D. et al, 2006, ApJ, 652, 1011
\bibitem{} Willingale R. et al, 2007, ApJ, 662, 1093
\bibitem{} Yost S. et al, 2007, ApJ, 657, 925
\bibitem{} Zhang B. et al, 2006, ApJ, 642, 354
\end{thebibliography}
\end{document}